\documentclass[twocolumn,english,conference]{IEEEtran}
\usepackage[T1]{fontenc}
\usepackage{amsmath}

\makeatletter
\usepackage[caption=false,font=footnotesize]{subfig}
\usepackage[utf8]{inputenc}
\usepackage{spconf}

\name{Sebastian Cammerer\IEEEauthorrefmark{1}, Benedikt Leible\IEEEauthorrefmark{1}, Matthias Stahl\IEEEauthorrefmark{1}, Jakob Hoydis\IEEEauthorrefmark{3}, Stephan ten Brink\IEEEauthorrefmark{1}}

\address{\IEEEauthorrefmark{1}Institute of Telecommunications, Pfaffenwaldring 47, University of Stuttgart, 70569 Stuttgart, Germany\\ Email: \{cammerer, leible, stahl, tenbrink\}@inue.uni-stuttgart.de\\
\IEEEauthorrefmark{3}Nokia Bell Labs, Route de Villejust, 91620 Nozay, France\\ Email: jakob.hoydis@nokia-bell-labs.com}

\usepackage{tikz}

\usetikzlibrary{dsp,chains,shapes}
\usepackage{graphicx}
\usepackage{pgfplots}

\usetikzlibrary{calc}
\usepackage{color}
\usepackage{xcolor}

\definecolor{mittelblau}{RGB}{0, 126, 198}
\definecolor{violettblau}{cmyk}{0.9, 0.6, 0, 0}
\definecolor{rot}{RGB}{238, 28 35}
\definecolor{apfelgruen}{RGB}{140, 198, 62}
\definecolor{gelb}{RGB}{1, 221, 0}
\definecolor{orange}{RGB}{244, 111, 33}
\definecolor{pink}{RGB}{237, 0, 140}
\definecolor{lila}{RGB}{128, 10, 145}
\definecolor{hellgrau}{RGB}{224, 224, 224}
\definecolor{mittelgrau}{RGB}{128, 128, 128}
\definecolor{dunkelgrau}{RGB}{80,80,80}
\definecolor{anthrazit}{RGB}{19, 31, 31}	

\tikzstyle{decision} = [diamond, draw, fill=blue!20,      text width=4.5em, text badly centered, node distance=3cm, inner sep=0pt] 
\tikzstyle{block} = [rectangle, draw, fill=blue!20,      text width=5em, text centered, rounded corners, minimum height=4em] 
\tikzstyle{line} = [draw, -latex'] 
\tikzstyle{cloud} = [draw, ellipse,fill=red!20, node distance=3cm,     minimum height=2em]

\colorlet{fncolor}{rot}
\newcommand{\frozennode}[2][fncolor]{\textcolor{#1}{#2}}

\@ifundefined{showcaptionsetup}{}{%
 \PassOptionsToPackage{caption=false}{subfig}}
\usepackage{subfig}
\makeatother

\usepackage{babel}
\begin{document}

\title{Combining Belief Propagation and Successive Cancellation List Decoding
of Polar Codes on a GPU Platform}
\maketitle
\begin{abstract}
The decoding performance of polar codes strongly depends on the decoding
algorithm used, while also the decoder throughput and its latency
mainly depend on the decoding algorithm. In this work, we implement
the powerful successive cancellation list (SCL) decoder on a GPU and
identify the bottlenecks of this algorithm with respect to parallel
computing and its difficulties. The inherent serial decoding property
of the SCL algorithm naturally limits the achievable speed-up gains
on GPUs when compared to CPU implementations. In order to increase
the decoding throughput, we use a hybrid decoding scheme based on
the belief propagation (BP) decoder, which can be intra- and inter-frame
parallelized. The proposed scheme combines excellent decoding performance
and high throughput within the signal-to-noise ratio (SNR) region
of interest. 
\end{abstract}

\section{Introduction}

Polar codes are proven to be capacity achieving under successive cancellation
(SC) decoding \cite{arikan_polar} for infinite block lengths. However,
for short lengths, SC decoding shows a weak performance compared to
state-of-the-art LDPC codes \cite{arikan_challenges_coding}. A major
breakthrough in polar decoding for short length codes was achieved
by Tal and Vardy with a successive cancellation list (SCL) decoder
\cite{tal/vardy_scl}. SCL decoding renders polar codes into a powerful
coding scheme whenever short block lengths are required \cite{niu_basic_concepts},
as for example for the internet of things (IoT) or very low latency
applications, both cornerstones of the upcoming 5G standard. However,
the issue of the high SCL decoding complexity needs to be solved before
polar codes can become competitive for practical applications. Besides
their excellent decoding performance, the code structure of polar
codes is inherently given by the concept of channel polarization \cite{arikan_polar},
making them attractive for upcoming communication standards. Additionally,
the code rate can be freely chosen by appropriately fixing a fraction
of frozen bits. 

One of the biggest current trends in the telecommunications industry
is virtualization with the goal of replacing specialized hardware
by software running on commodity servers \cite{rost2015benefits}.
However, performance critical software components, e.g., signal processing
for the physical layer, require the use of hardware accelerators (GPUs,
FPGAs) to satisfy the strict latency and throughput requirements of
next generation communication systems. Unfortunately, not all algorithms
or processing steps benefit from acceleration because they are either
difficult to parallelize or the performance gains are eaten up by
the overhead related to memory access. One of the most processing-resource
consuming components of the physical layer is channel decoding \cite{nikaein2015processing}.
Thus, an efficient hardware-accelerated software implementation of
a decoder is of utmost importance for virtualized communication systems. 

In this work we focus on graphical processing unit (GPU) implementations
using the \textit{NVIDIA Compute Unified Device Architecture (CUDA)}
Framework \cite{nvidia_cuda}. For low-density parity-check (LDPC)
codes high throughput gains were observed for LDPC belief propagation
(BP) decoding \cite{Falcao_ldpc_gpu}, where massive parallelization
can be done straightforwardly via parallel node updates. A similar
BP algorithm can also be used for polar code decoding \cite{BP_arikan}
and a correspondingly high throughput gain was observed in \cite{reddy_polar_bp_gpu}
on a GPU. Nonetheless, the implementation of the SCL algorithm for
parallel computing requires more efforts, as this algorithm uses recursive
updates. According to our knowledge, no GPU-based SCL decoder is proposed
in the literature. Very recently an implementation of the \textit{fast
simplified successive cancellation} (fast SSC) decoder was shown in
\cite{giard_gpu_decoder}, but no remarkable speed-up could be observed
when compared to CPU implementations\footnote{Remark: Several optimized SC(L) algorithms exist, mostly based on
pruning/unrolling of the decoding graph. We stick to the plain SCL
decoder, as this seems to be the most flexible algorithm (i.e., in
terms of the code rate and variable frozen bit positions). The hybrid
scheme works as well for a SSC(L) implementation of the SCL decoder.
In particular for research applications, full SCL decoding is required
without any further simplifications.}. Although, the overall performance is impressive, one needs to keep
in mind that a quantization with 8-bit per value (32-bit floating
point precision in our implementation) and no list decoding is assumed,
i.e., the decoding complexity increases (linearly) with the list size
$L$ which, typically, is set to $L=32$.

In this paper, we identify the bottlenecks of such an implementation
for parallel computing. It turns out that the SCL algorithm itself
has a limited potential for GPU implementations. Therefore, we propose
an alternative approach, where we combine the SCL decoder together
with a belief propagation (BP) decoder. This concept combines the
best of both worlds, i.e., good error correction capability and high
throughputs. The authors in \cite{yuan_hybrid_sc} propose a combination
of SC and BP decoding and also show the possibility of using the same
hardware for both algorithms. This idea is also similar to the proposed
adaptive SCL decoder \cite{Li_adaptive_list}, which increases the
list size whenever decoding fails. However, these algorithms increase
the decoding latency, since two decoding steps are required, although
the average latency improves. We examine the decoding latency for
a given signal-to-noise ratio (SNR) and for different code rates.

All simulations are performed on an \textit{Intel i7-4790K CPU @ 4.00GHz}
and a \textit{NVIDIA GTX 980 Ti} (with \textit{6 GB GDDR5} Memory).
We also provide our decoder online as a webdemo \cite{webdemo}, where
variable block lengths and arbitrary frozen bit patterns can be simulated
on-the-fly (in real-time) on our servers.

\section{Polar Codes}

\begin{figure}[t]
\begin{centering}
\begin{center}
\begin{center}
\begin{tikzpicture}[y=0.55cm]
\node[dspnodefull,minimum size=1mm] (u1) at (0, 1) {$u_6$};
\node[dspnodefull,minimum size=1mm] (u2) at (0, 0) {$u_7$};
\node[dspnodefull,minimum size=1mm] (temp) at (1, 0) {};
\node[dspadder](xor) at (1,1) {};
\node[dspnodefull,minimum size=1mm] (x1) at (2, 1) {};
\node[dspnodefull,minimum size=1mm] (x2) at (2, 0) {};
\draw[line width = 0.5mm](u1)--node[above] {}(xor);
\draw[line width = 0.5mm](temp)--(xor);
\draw[line width = 0.5mm](xor)--node[above] {}(x1);
\draw[line width = 0.5mm](temp)--node[above] {}(x2);
\draw[line width = 0.5mm](u2)--node[above] {}(temp);

\node[dspnodefull,minimum size=1mm,color=fncolor] (u12) at (0, 3) {\frozennode{$u_4$}};
\node[dspnodefull,minimum size=1mm] (u22) at (0, 2) {$u_5$};
\node[dspnodefull,minimum size=1mm] (temp2) at (1, 2) {};
\node[dspadder](xor2) at (1,3) {};
\node[dspnodefull,minimum size=1mm] (x12) at (2, 3) {};
\node[dspnodefull,minimum size=1mm] (x22) at (2, 2) {};
\draw[line width = 0.5mm,color=fncolor](u12)--node[above] {}(xor2);
\draw[line width = 0.5mm](temp2)--(xor2);
\draw[line width = 0.5mm](xor2)--node[above] {}(x12);
\draw[line width = 0.5mm](temp2)--node[above] {}(x22);
\draw[line width = 0.5mm](u22)--node[above] {}(temp2);

\node[] (u13) at (-0.5,3.5) {$\boldsymbol{u}$};
\node[] (u13) at (7.5,3.5) {$\boldsymbol{x}$};

\node[dspnodefull,minimum size=1mm,color=fncolor] (u13) at (0, 5) {\frozennode{$u_2$}};
\node[dspnodefull,minimum size=1mm] (u23) at (0, 4) {$u_3$};
\node[dspnodefull,minimum size=1mm] (temp3) at (1, 4) {};
\node[dspadder](xor3) at (1,5) {};
\node[dspnodefull,minimum size=1mm] (x13) at (2, 5) {};
\node[dspnodefull,minimum size=1mm] (x23) at (2, 4) {};
\draw[line width = 0.5mm,color=fncolor](u13)--node[above] {}(xor3);
\draw[line width = 0.5mm](temp3)--(xor3);
\draw[line width = 0.5mm](xor3)--node[above] {}(x13);
\draw[line width = 0.5mm](temp3)--node[above] {}(x23);
\draw[line width = 0.5mm](u23)--node[above] {}(temp3);

\node[dspnodefull,minimum size=1mm,color=fncolor] (u14) at (0, 7) {\frozennode{$u_0$}};
\node[dspnodefull,minimum size=1mm,color=fncolor] (u24) at (0, 6) {\frozennode{$u_1$}};
\node[dspnodefull,minimum size=1mm,color=fncolor] (temp4) at (1, 6) {};
\node[dspadder,color=fncolor](xor4) at (1,7) {};
\node[dspnodefull,minimum size=1mm,color=fncolor] (x14) at (2, 7) {};
\node[dspnodefull,minimum size=1mm,color=fncolor] (x24) at (2, 6) {};
\draw[line width = 0.5mm,color=fncolor](u14)--node[above] {}(xor4);
\draw[line width = 0.5mm,color=fncolor](temp4)--(xor4);
\draw[line width = 0.5mm,color=fncolor](xor4)--node[above] {}(x14);
\draw[line width = 0.5mm,color=fncolor](temp4)--node[above] {}(x24);
\draw[line width = 0.5mm,color=fncolor](u24)--node[above] {}(temp4);

\node[dspnodefull,minimum size=1mm] (temp5) at (3, 0) {};
\node[dspadder](xor5) at (3,2) {};
\node[dspnodefull,minimum size=1mm] (y22) at (4, 0) {};
\node[dspnodefull,minimum size=1mm] (y42) at (4, 2) {};
\draw[line width = 0.5mm](x22)--node[above] {}(xor5);
\draw[line width = 0.5mm](temp5)--(xor5);
\draw[line width = 0.5mm](xor5)--node[above] {}(y42);
\draw[line width = 0.5mm](temp5)--node[above] {}(y22);
\draw[line width = 0.5mm](x2)--node[above] {}(temp5);

\node[dspnodefull,minimum size=1mm] (temp6) at (2.5, 1) {};
\node[dspadder](xor6) at (2.5,3) {};
\node[dspnodefull,minimum size=1mm] (y12) at (4, 1) {};
\node[dspnodefull,minimum size=1mm] (y32) at (4, 3) {};
\draw[line width = 0.5mm](x12)--node[above] {}(xor6);
\draw[line width = 0.5mm](temp6)--(xor6);
\draw[line width = 0.5mm](xor6)--node[above] {}(y32);
\draw[line width = 0.5mm](temp6)--node[above] {}(y12);
\draw[line width = 0.5mm](x1)--node[above] {}(temp6);

\node[dspnodefull,minimum size=1mm] (temp7) at (3, 4) {};
\node[dspadder](xor7) at (3,6) {};
\node[dspnodefull,minimum size=1mm] (y6) at (4, 4) {};
\node[dspnodefull,minimum size=1mm] (y8) at (4, 6) {};
\draw[line width = 0.5mm,color=fncolor](x24)--node[above] {}(xor7);
\draw[line width = 0.5mm](temp7)--(xor7);
\draw[line width = 0.5mm](xor7)--node[above] {}(y8);
\draw[line width = 0.5mm](temp7)--node[above] {}(y6);
\draw[line width = 0.5mm](x23)--node[above] {}(temp7);

\node[dspnodefull,minimum size=1mm] (temp8) at (2.5, 5) {};
\node[dspadder](xor8) at (2.5,7) {};
\node[dspnodefull,minimum size=1mm] (y5) at (4, 5) {};
\node[dspnodefull,minimum size=1mm] (y7) at (4, 7) {};
\draw[line width = 0.5mm,color=fncolor](x14)--node[above] {}(xor8);
\draw[line width = 0.5mm](temp8)--(xor8);
\draw[line width = 0.5mm](xor8)--node[above] {}(y7);
\draw[line width = 0.5mm](temp8)--node[above] {}(y5);
\draw[line width = 0.5mm](x13)--node[above] {}(temp8);

\node[dspnodefull,minimum size=1mm] (temp9) at (4.5, 3) {};
\node[dspadder](xor9) at (4.5,7) {};
\node[dspnodefull,minimum size=1mm] (z1) at (7, 3) {$x_4$};
\node[dspnodefull,minimum size=1mm] (z5) at (7, 7) {$x_0$};
\draw[line width = 0.5mm](y7)--node[above] {}(xor9);
\draw[line width = 0.5mm](temp9)--(xor9);
\draw[line width = 0.5mm](xor9)--node[above] {}(z5);
\draw[line width = 0.5mm](temp9)--node[above] {}(z1);
\draw[line width = 0.5mm](y32)--node[above] {}(temp9);

\node[dspnodefull,minimum size=1mm] (temp10) at (5, 2) {};
\node[dspadder](xor10) at (5,6) {};
\node[dspnodefull,minimum size=1mm] (z2) at (7, 2) {$x_5$};
\node[dspnodefull,minimum size=1mm] (z6) at (7, 6) {$x_1$};
\draw[line width = 0.5mm](y8)--node[above] {}(xor10);
\draw[line width = 0.5mm](temp10)--(xor10);
\draw[line width = 0.5mm](xor10)--node[above] {}(z6);
\draw[line width = 0.5mm](temp10)--node[above] {}(z2);
\draw[line width = 0.5mm](y42)--node[above] {}(temp10);

\node[dspnodefull,minimum size=1mm] (temp11) at (5.5, 1) {};
\node[dspadder](xor11) at (5.5,5) {};
\node[dspnodefull,minimum size=1mm] (z3) at (7, 1) {$x_6$};
\node[dspnodefull,minimum size=1mm] (z7) at (7, 5) {$x_2$};
\draw[line width = 0.5mm](y5)--node[above] {}(xor11);
\draw[line width = 0.5mm](temp11)--(xor11);
\draw[line width = 0.5mm](xor11)--node[above] {}(z7);
\draw[line width = 0.5mm](temp11)--node[above] {}(z3);
\draw[line width = 0.5mm](y12)--node[above] {}(temp11);

\node[dspnodefull,minimum size=1mm] (temp12) at (6, 0) {};
\node[dspadder](xor12) at (6,4) {};
\node[dspnodefull,minimum size=1mm] (z4) at (7, 0) {$x_7$};
\node[dspnodefull,minimum size=1mm] (z8) at (7, 4) {$x_3$};
\draw[line width = 0.5mm](y6)--node[above] {}(xor12);
\draw[line width = 0.5mm](temp12)--(xor12);
\draw[line width = 0.5mm](xor12)--node[above] {}(z8);
\draw[line width = 0.5mm](temp12)--node[above] {}(z4);
\draw[line width = 0.5mm](y22)--node[above] {}(temp12);
\end{tikzpicture}
\par\end{center}
\par\end{center}
\par\end{centering}

\begin{centering}
\caption{Polar encoder graph for $N=8$; red color indicates the frozen bit
positions.\label{fig:encoding_circuit}}

\par\end{centering}

\vspace{-0.5cm}
\end{figure}
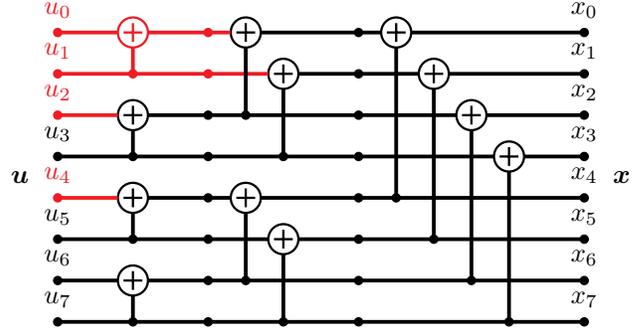

The concept of channel polarization \cite{arikan_polar} transforms
$N$ independent channels into polarized channels by channel combining
and splitting, i.e., a set of more reliable and a set of less reliable
channels can be observed. The $k$ most reliable channels are now
used to transmit information bits, while the other $N-k$ bit-channels
are frozen which, w.l.o.g., are set to zero. For a given code rate
$R=k/N$, the $k$ information bits are mapped onto the $k$ non-frozen
positions of ${\boldsymbol{u}}_{N}$, all other positions are frozen.
The encoding process (polar transformation) can be simply described
by a generator matrix ${\boldsymbol{\mathrm{G}}=\boldsymbol{\mathrm{F}}^{\otimes n}}$,
where $\boldsymbol{\mathrm{F}}^{\otimes n}$ denotes the $n^{th}$
Kronecker power of $\boldsymbol{\mathrm{F}}=\begin{bmatrix}1 & 0\\
1 & 1
\end{bmatrix}$. The transmitted codeword is ${\boldsymbol{x}}={\boldsymbol{u}\cdot\boldsymbol{\mathrm{G}}}$.
As can be seen in Fig. \ref{fig:encoding_circuit}, the resulting
encoding circuit has complexity $\mathcal{O}\mbox{\ensuremath{\left(N\cdot\log N\right)}}$.

Since the selection of the frozen positions is not the main topic
of the paper, we consider a given (arbitrary) frozen bit vector $\mathbf{{f}}$
throughout this paper. For more details, we refer the interested reader
to \cite{tal/vardy_construct_pcodes}.

\subsection{SC-List Decoding on GPUs}

The SCL decoder utilizes the bitwise-serial decoding algorithm of
the SC decoder \cite{arikan_polar} and adds a list, holding up to
$L$ of the most probable paths for the estimated codeword $\boldsymbol{\hat{x}}$
of length $N=2^{n}$, resulting in a overall decoding complexity of
$\mathcal{O}\mbox{\ensuremath{\left(L\cdot N\cdot\log(N)\right)}}$
\cite{tal/vardy_scl}. Every entry $\ell\in\left\{ 1,2,\ldots,2L\right\} $
of the list is updated for every bit decision $\hat{u}_{\ell,i}$,
with $i$ being the bit position of the estimated message bit and
$\boldsymbol{\hat{u}}_{\ell}$ being the estimated message vector
for list entry $\ell$. List updates include sorting the branching
options for each path by a metric $m_{\ell,i}$ giving the probability
that the corresponding path is representing the correct estimate of
the transmitted codeword. Hence, paths can get discarded and replaced
by more promising candidates, which results in the need to duplicate
data for newly listed branches frequently. The SCL decoding algorithm
can be split up into two parts for each decision $\hat{u}_{\ell,i}$
that is made concerning an unfrozen bit. Log-likelihood ratios (LLR)
are given by $LLR\left(y_{j}\right)=\mathrm{ln}\left(\frac{P(y_{j}\left|x_{j}=0)\right.}{P(y_{j}\left|x_{j}=1)\right.}\right)$,
where $P(y_{j}\left|x_{j})\right.$ denotes the conditional probability
that the channel output $y_{j}$ is received while the codeword element
$x_{j}$ was transmitted. First an $LLR\left(\hat{u}_{i,\ell}\right)=\mathrm{ln}\left(\frac{P(\boldsymbol{y},\hat{\boldsymbol{u}}_{0,\ell}^{i-1}\left|u_{i}=0)\right.}{P(\boldsymbol{y},\hat{\boldsymbol{u}}_{0,\ell}^{i-1}\left|u_{i}=1)\right.}\right)$
is calculated from the received codewords $\boldsymbol{y}$ and the
previously decided bits $\hat{\boldsymbol{u}}_{0,\ell}^{i-1}$ of
the respective list entry $\ell$. Subsequently, the calculated values
are used to decide which of the maximal $2L$ branching options are
kept in the list for the next decoding phase. Additionally, a cyclic
redundancy check (CRC) can be added, aiding the selection of the estimated
codeword after the last decision step \cite{tal/vardy_scl}. This
further improves the decoding capabilities of the SCL decoder and
results in a negligibly smaller code rate $R_{\mathrm{crc}}=\frac{k-c}{N}$,
with $c$ being the number of CRC parity bits.

For the implementation, the CUDA framework by NVIDIA is used, which
enables the use of commodity graphics hardware for single instruction
multiple data (SIMD) parallel programming \cite{nvidia_cuda}. Due
to high data dependencies between the bitwise estimation steps, the
algorithm of the SCL decoder is mostly serial by design. There seems
to be no straightforward approach for a massively parallel implementation
of this particular decoder, which results in impractical throughputs
for a parallel implementation of the standard SCL decoder. Even though
the bitwise decoding steps cannot be parallelized due to data dependencies,
many of the calculations per step can be parallelized. Also, multiple
list places can be processed in parallel but have to be synchronized
for every decision on an unfrozen bit.

For sorting the $2L$ candidate paths by their probabilities, a parallel
bitonic sort \cite{peters2011fast} was implemented. Additionally,
we propose a ``pruned pseudo sort'' to determine the $L$ smallest
path metrics in parallel. This pseudo sort algorithm calculates $d_{\ell,i}=\underset{j=1}{\overset{2L}{\sum}}\mathrm{H}(m_{\ell,i}-m_{j,i})$
for all $\ell\in\left\{ 1,2,\ldots,2L\right\} $, with $\mathrm{H}(x)=\begin{cases}
0, & x<0\\
1, & x\geq0
\end{cases}$ being the Heaviside function, and discards all candidates with $d_{\ell,i}>L$.

Further, the \textit{decision aiding mechanism}, described in \cite{Li_decision_aided},
was implemented to reduce the cases in which the decoder has to execute
the sort and duplicate methods, which represent the dominant bottleneck
reducing the achievable throughput. Instead, the more reliable (information)
bit positions are estimated without branching the list (according
to their Bhattacharya parameter \cite{arikan_polar}). For a carefully
selected amount of bits, the decoding performance does not degrade,
while the throughput strongly increases \cite{Li_decision_aided}.

\subsection{Belief Propagation Decoding of Polar Codes}

\begin{figure}[t]
\centering{}\begin{center}
\begin{tikzpicture}[y=0.6cm]
\node[] (p1) at (0.4, -1.8) {}; 
\node[] (p2) at (1.6, -1.8) {}; 
\node[] (p3) at (0.4, 2) {}; 
\node[] (p4) at (1.6, 2) {};
\filldraw [fill=gray!20!white,draw=gray,dashed,line width = 0.3mm] (p1) rectangle (p4);
\node[scale=1.5,gray] (p5) at (1, 2.8) {PE};
\node[] (u1) at (-2, 1) {};
\node[] (u2) at (-2, -1) {};
\node[dspnodefull,minimum size=0.2mm] (temp) at (1, -1) {};
\node[dspadder](xor) at (1,1) {};
\node[] (x1) at (4, 1) {};
\node[] (x2) at (4, -1) {};
\draw[line width = 0.5mm](u1)--node[above=0.1cm] {$R_{\mathrm{in,1}}$}(xor);
\draw[line width = 0.5mm](u1)--node[below=0.1cm] {$L_{\mathrm{out,1}}$}(xor);
\draw[dspconn,line width = 0.25mm]      (-1.2,1.25) --node[above] {} (0.3,1.25);
\draw[dspconn,line width = 0.25mm]      (0.3,0.75) --node[above] {} (-1.2,0.75); \draw[line width = 0.5mm](temp)--(xor);
\draw[line width = 0.5mm](xor)--node[above=0.1cm] {$R_{\mathrm{out,1}}$}(x1);
\draw[line width = 0.5mm](xor)--node[below=0.1cm] {$L_{\mathrm{in,1}}$}(x1);
\node[] (shiftxy) at (2.95, 2) {};
\draw[dspconn,line width = 0.25mm]      ($(shiftxy)+(-1.2,-0.75)$) --node[above] {} ($(shiftxy)+(0.3,-0.75)$);
\draw[dspconn,line width = 0.25mm]      ($(shiftxy)+(0.3,-1.25)$) --node[above] {} ($(shiftxy)+(-1.2,-1.25)$);
\draw[line width = 0.5mm](temp)--node[above=0.1cm] {$R_{\mathrm{out,2}}$}(x2);
\draw[line width = 0.5mm](temp)--node[below=0.1cm] {$L_{\mathrm{in,2}}$}(x2);
\node[] (shiftx) at (2.95, 0) {};
\draw[dspconn,line width = 0.25mm]      ($(shiftx)+(-1.2,-0.75)$) --node[above] {} ($(shiftx)+(0.3,-0.75)$);
\draw[dspconn,line width = 0.25mm]      ($(shiftx)+(0.3,-1.25)$) --node[above] {} ($(shiftx)+(-1.2,-1.25)$);
\draw[line width = 0.5mm](u2)--node[above=0.1cm] {$R_{\mathrm{in,2}}$}(temp);
\draw[line width = 0.5mm](u2)--node[below=0.1cm] {$L_{\mathrm{out,2}}$}(temp);
\draw[dspconn,line width = 0.25mm]      (-1.2,-0.75) --node[above=0.1cm] {} (0.3,-0.75);
\draw[dspconn,line width = 0.25mm]      (0.3,-1.25) --node[above=0.1cm] {} (-1.2,-1.25);

\node[dspnodefull]  at (-1.85,1) {};
\node[dspnodefull]  at (-1.85,-1) {};
\node[dspnodefull]  at (3.85,1) {};
\node[dspnodefull]  at (3.85,-1) {};
\end{tikzpicture}
\par\end{center}\vspace{-0.6cm}
\caption{Single processing element (PE) of the BP decoder.\label{fig:bp_dec}}
\vspace{-0.3cm}
\end{figure}
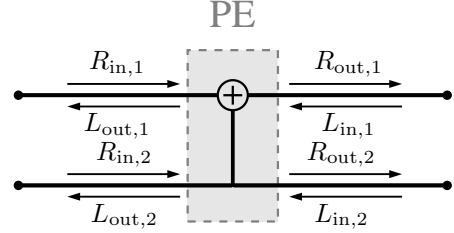

BP decoding of polar codes is a message passing algorithm based on
the encoding scheme shown in Fig. \ref{fig:encoding_circuit} with
decoding complexity $\mathcal{O}\mbox{\ensuremath{\left(n\cdot\log n\right)}}$.
The transmitted codeword $\boldsymbol{\hat{x}}$ and the message $\boldsymbol{\hat{u}}$
can be both estimated simultaneously. There are $n+1$ stages with
$N$ nodes per stage. Each stage consists of $N/2$ processing elements
(PEs) which are iteratively passing messages to adjacent nodes. Each
PE connects $4$ nodes in $2$ consecutive stages as shown in Fig.~\ref{fig:bp_dec},
the input/output relation of the PEs is the same as in Fig. \ref{fig:encoding_circuit}.
All messages are calculated in the LLR domain. One decoding iteration
consists of two steps, where the soft messages are updated at each
PE (until reaching a maximum number of iterations) as follows:
\begin{enumerate}
\item Update left-to-right messages, called \textbf{$\mathbf{R}$}-messages,
for stages $2,...,n+1$\begin{align*} R_{\mathrm{out},1} &=\mathrm{g}(R_{\mathrm{in},1},L_{\mathrm{in},2}+R_{\mathrm{in},2}) \\ R_{\mathrm{out},2} &=\mathrm{g}(R_{\mathrm{in},1},L_{\mathrm{in},1})+R_{\mathrm{in},2}  \end{align*}
\item Update right-to-left messages, called $\mathbf{L}$-messages, for
stages $n,...,1$ \begin{align*} L_{\mathrm{out},1} &=\mathrm{g}(L_{\mathrm{in},1},L_{\mathrm{in},2}+R_{\mathrm{in},2}) \\ L_{\mathrm{out},2}&=\mathrm{g}(R_{\mathrm{in},1},L_{\mathrm{in},1})+L_{\mathrm{in},2} \end{align*} 
\end{enumerate}
\noindent where $\mathrm{g}\left(a,b\right)=\mathrm{ln}\left(\frac{1+\mathrm{e}^{a+b}}{\mathrm{e}^{a}+\mathrm{e}^{b}}\right)$.
For the $\mathrm{g}\left(\cdot\right)$-function, a min-approximation
$\mathrm{g}(a,b)=\mathrm{sign}(a)\cdot\mathrm{sign}(b)\cdot\min(\left|a\right|,\left|b\right|)$
can be used which is more suitable for hardware implementation \cite{sung_bp_fpga}.
An advantage of GPU computing is the availability of many floating-point
units (FPU). Thus, we apply the exact node update-equations with clipping
the absolute values of the LLR values to $LLR_{max}=20$ to ensure
numerical stability.

\noindent 
We initialize $L_{i,n+1}=LLR(y_{i})$ and $R_{i,1}=LLR_{max}\cdot f_{i}$.
The final output of the decoder is \begin{equation*}\label{eq:L_u_x_i}\begin{split}LLR\left(\hat{u}_{i}\right) & =  L_{i,1}+R_{i,1} \\ LLR\left(\hat{x}_{i}\right) & =  L_{i,n+1}+R_{i,n+1}.  \end{split} \end{equation*} 

\noindent A hard-decision gives the estimated bit-vectors, while for
other applications such as a combined channel-detection, soft-values
may be required and can be easily forwarded. Additionally, stopping
conditions \cite{yuan_stopping_cond} exist, and thus, the decoding
can be done within a few iterations for most of the frames (in the
high SNR region) as soon as ${\boldsymbol{\hat{x}}}={\boldsymbol{\hat{u}}\cdot\boldsymbol{\mathrm{G}}}$.
For this specific setup, a simple CRC on $\boldsymbol{\hat{u}}$ can
replace the more complex re-encoding step, i.e., decoding is stopped
whenever the CRC of $\boldsymbol{\hat{u}}$ is fulfilled (or a maximum
of iterations is performed). 

\noindent We apply intra- and inter-frame parallelism, i.e., several
codewords are decoded in parallel, where each single thread implements
one PE. Nevertheless, synchronization after each stage is required.

\section{Combining BP and SCL Decoding }

\begin{figure}[t]
\begin{centering}
\vspace{0.15cm}
\resizebox{0.45\textwidth}{!}{
\begin{tikzpicture}[node distance = 2cm,auto]     
\definecolor{graycolor}{HTML}{B7DDE8}
\node [cloud,fill=graycolor] (start) {\ \ \ Start\ \ \ };
\node [decision, below of=start] (checkbuff) {enough buffer space?}; 
\node [block, above left= 0.4cm and 0.7cm of checkbuff,node distance = 3cm,fill={rgb:red,0;green,5;blue,2}] (startbpsh2) {start BP decoder};
\node [block, above left= 0.3cm and 0.6cm of checkbuff,node distance = 3cm,fill={rgb:red,0;green,5;blue,2}] (startbpsh3) {start BP decoder};
\node [block, above left= 0.20cm and 0.5cm of checkbuff,node distance = 3cm,fill={rgb:red,0;green,5;blue,2}] (startbp) {start BP decoder};    
\node [decision, below of=startbp,fill={rgb:red,0;green,5;blue,2}] (bpfinished) {BP finished CWs?};
\node [decision, left of= bpfinished,fill={rgb:red,0;green,5;blue,2},node distance=0.15cm] (bpfinishedsh1) {BP finished CWs?};     
\node [decision, below of=startbp,fill={rgb:red,0;green,5;blue,2}] (bpfinishedover) {BP finished CWs?};
\node [decision, left of= bpfinished,fill={rgb:red,0;green,5;blue,2},node distance=-0.15cm] (bpfinishedsh02) {BP finished CWs?};      
\node [block, below of= bpfinished,node distance = 3cm,fill={rgb:red,0;green,5;blue,2}] (copytobuff) {copy failed CWs to buffer};     
\node [block, below right= -1.33cm and -1.9cm of copytobuff,fill={rgb:red,0;green,5;blue,2}] (copytobuffsh1) {copy failed CWs to buffer};
\node [block, below right= -1.33cm and -1.9cm of copytobuffsh1,fill={rgb:red,0;green,5;blue,2}] (copytobuffsh2) {copy failed CWs to buffer};
\node [decision, below of=checkbuff] (checksclbuff) {buffer not empty \& SCLs idle?};

\node [block, block, below right= 0.025cm and 0.50cm of checksclbuff,fill={rgb:red,0;green,5;blue,2},node distance = 3cm] (startscl) {start idle SCL decoders}; \node [block, below right= -1.33cm and -1.9cm of startscl,fill={rgb:red,0;green,5;blue,2}] (startsclsh1) {start idle SCL decoders}; \node [block, below right= -1.33cm and -1.9cm of startsclsh1,fill={rgb:red,0;green,5;blue,2}] (startsclsh2) {start idle SCL decoders}; 
\node [decision, above of=startscl] (sclfinish) {SCL decoder finished?};
\node [decision, above of=sclfinish] (breakcond) {all CWs decoded?};
\node [cloud,, above right= 0.45cm and 0.9cm of breakcond,node distance = 2.55cm,fill=graycolor] (stop) {\ \ \ Stop\ \ \ };     
\node [block, below right= 0.025cm and 0.70cm of breakcond] (counterror) {free buffers};

\path [line] (start) -> (checkbuff);     
\path [line] (checkbuff) -> node {yes}(startbp);     
\path [line] (startbp) -> (bpfinished);
\path [line] (bpfinished) -> node {yes}(copytobuff);
\path [line] (checkbuff) -> node {no}(checksclbuff);   
\path [line] (bpfinishedsh02) -> node {no}(checksclbuff);   
\path [line] (checksclbuff) -> node {yes}(startscl);
\path [line] (copytobuff) -> (checksclbuff);
\path [line] (breakcond) -> node {yes}(stop);
\path [line] (sclfinish) -> node {yes}(counterror);     
\path [line] (startscl) -> (sclfinish);
\path [line] (counterror) -> (breakcond);
\path [line] (checksclbuff) -> node {no}(sclfinish);     
\path [line] (sclfinish) -> node {no}(breakcond);     
\path [line] (breakcond) -> node {no}(checkbuff);

\end{tikzpicture}
}
\par\end{centering}

\caption{Scheduling mechanism of the hybrid decoder (grey nodes: CPU, green
nodes: GPU). \label{fig:block_scheduling}}
\end{figure}
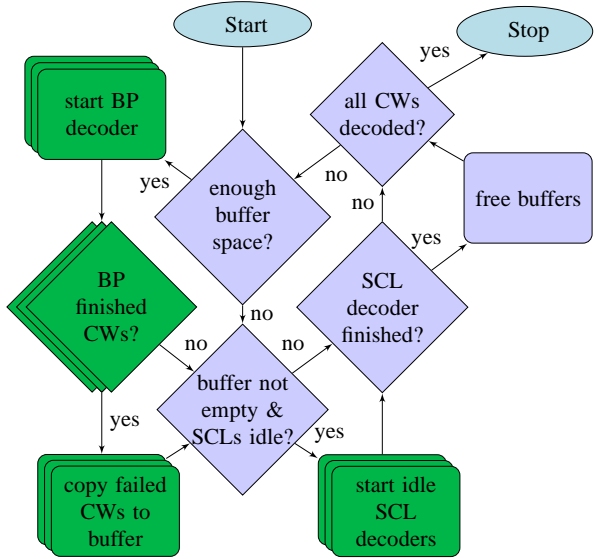

The bit-error-rate (BER) performance of polar codes under SCL decoding
is better than that under BP decoding \cite{niu_basic_concepts}.
However, in terms of suitability of parallelization, the BP decoder
shows a higher potential because all bits can be calculated in parallel
while the latency can be decreased. This is shown in Fig. \ref{fig:block_scheduling}
and Fig. \ref{fig:dec_latency}, respectively.

Whenever the CRC after the maximum number of BP iterations $i_{\mathrm{max}}$
does not hold, the codeword is forwarded to an additional SCL decoding
step. It turns out that only $i_{\mathrm{max}}=50$ BP iterations
are sufficient, otherwise each decoding failure blocks GPU ressources
for a long time.\footnote{Remark: The overall BER performance does not (significantly) depend
on $i_{\mathrm{max}}$, since we assume that SCL decoding can decode
(practically) all noisy codewords anyway, which could be decoded under
BP decoding.} 

SCL decoding could be performed on the CPU as well, however, the required
data transfer\textit{ }produces additional latency overhead and is
thus not advisable.

\section{Decoding Performance}

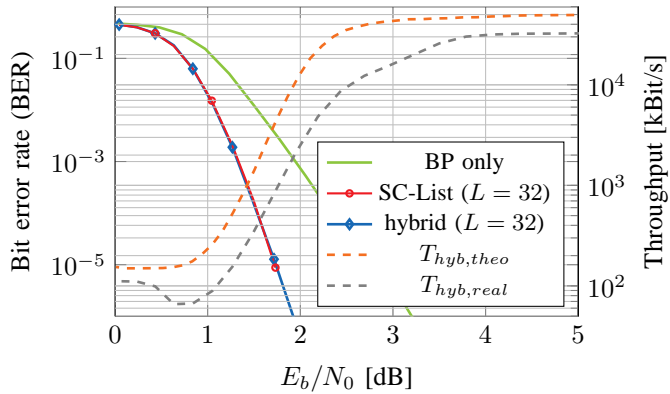
\begin{figure}[t]
\begin{tikzpicture}

\pgfplotsset{
height=5.7cm,width=0.9\columnwidth,
xmin=0, xmax=5,
xlabel={$E_b/N_0$ [dB]},
grid=both,
legend pos= south east,
legend style={font=\small},}

\begin{semilogyaxis}[
ymin=1e-6,ymax=1,
ylabel={Bit error rate (BER)}
]
\addplot [apfelgruen,line width=1.0pt]table {results/run1.dat};
\label{plot_1};
\addplot [violettblau,mark=diamond,mark repeat=2, mark size = 1.8pt,line width=1.0pt]table {results/hybrid_new.dat};
\label{plot_4};
\addplot [rot,mark=o,mark repeat=3,mark size=1.2pt,line width=0.8pt]table {results/scl.dat};
\label{plot_3};
\end{semilogyaxis}

\begin{semilogyaxis}[
axis y line*=right,
axis x line=none,
ylabel near ticks,
yticklabel pos=right,
ymin=5e1,ymax=6e4,
ylabel={Throughput [kBit/s]},]
\addlegendimage{/pgfplots/refstyle=plot_1}\addlegendentry{BP only}
\addlegendimage{/pgfplots/refstyle=plot_3}\addlegendentry{SC-List ($L=32$)}
\addlegendimage{/pgfplots/refstyle=plot_4}\addlegendentry{hybrid ($L=32$)}

\addplot [orange,dashed,line width=1.0pt]table {results/thru_theo.dat};
\addlegendentry{$T_{hyb,theo}$};
\addplot [mittelgrau,dashed,line width=1.0pt]table {results/thru_new.dat};
\addlegendentry{$T_{hyb,real}$};
\end{semilogyaxis}

\end{tikzpicture}

\caption{Hybrid decoder throughput $T_{hyb,real}$ measurement, approximated
throughput $T_{hyb,theo}$ and BER performance of the BP, the SCL
and the hybrid decoder with $N=4096$ and $R=0.5$. \label{fig:dec_throughput}}
\end{figure}

For a given BP frame-error-rate (FER) $\gamma_{BP,FER}$ and a (information
bit) throughput $T_{BP}$ and $T_{SCL}$ for the BP and SCL decoder,
respectively, the overall throughput can be approximated (without
kernel-call overhead) as 
\begin{equation}
T_{hyb,theo}=\frac{T_{BP}\cdot T_{SCL}}{T_{SCL}+\gamma_{BP,FER}\cdot T_{BP}}.\label{eq:throughput}
\end{equation}
The measured throughput is depicted in Fig. \ref{fig:dec_throughput},
the gap between theoretical and measured throughput can be explained
by the kernel-call overhead, which is not considered in (\ref{eq:throughput}).
A maximum decoding throughput of $34$ Mbit/s can be achieved for
$N=4096$, $L=32$ and $R=0.5$. Additionally, it can be seen that
the BER does not differ from the SCL curve. 

We need to emphasize that the operation point of channel codes is
typically not in the high BER region (as no reliable communication
is possible) and thus high speed-up gains are observed in practice.
At least for low (and intermediate) BER applications, e.g., error-floor
analysis which requires a lot of simulation time, this proposal shows
a huge improvement by a factor of 100 and more, when compared to the
non-hybrid implementation of the SCL decoder.

\subsection{Latency}

\begin{figure}[t]
\begin{centering}
\subfloat[average decoding latency]{\centering{}
\begin{tikzpicture}

\pgfplotsset{
height=3.8cm,width=\columnwidth,
xmin=0, xmax=5,
xlabel={$E_b/N_0$ [dB]},
grid=both,
legend pos= north east,
legend style={font=\small}}

\begin{semilogyaxis}[
ymin=1e-3,ymax=1e1,
ylabel={avg. latency [s]}
]

\addplot [mittelblau,mark=o,mark repeat=3, mark size = 1.2pt,line width=0.8pt]table {results/latency_avg_bp.dat};
\addlegendentry{$L_{BP}$};
\addplot [apfelgruen,mark=x,mark repeat=3,line width=1.0pt]table {results/latency_avg_scl.dat};
\addlegendentry{$L_{SCL}$};
\addplot [rot,mark=diamond,mark repeat=2,mark size=1.8pt,line width=1.0pt]table {results/latency_avg_hybrid.dat};
\addlegendentry{$L_{hyb}$};

\end{semilogyaxis}

\end{tikzpicture}
}\vspace{-0.3cm}
\\
\subfloat[worst-case decoding latency]{\centering{}\begin{tikzpicture}

\pgfplotsset{
height=3.8cm,width=\columnwidth,
xmin=0, xmax=5,
xlabel={$E_b/N_0$ [dB]},
grid=both,
legend pos= north east,
legend style={font=\small}}

\begin{semilogyaxis}[
ymin=1e-3,ymax=1e1,
ylabel={max. latency [s]}
]

\addplot [mittelblau,mark=o,mark repeat=3, mark size = 1.2pt,line width=0.8pt,restrict x to domain=0:4.3]table {results/latency_max_bp.dat};
\addlegendentry{$L_{BP}$};
\addplot [apfelgruen,mark=x,mark repeat=3,line width=1.0pt]table {results/latency_max_scl.dat};
\addlegendentry{$L_{SCL}$};
\addplot [rot,mark=diamond,mark repeat=2,mark size=1.8pt,line width=1.0pt,restrict x to domain=0:4.3]table {results/latency_max_hybrid.dat};
\addlegendentry{$L_{hyb}$};

\end{semilogyaxis}

\end{tikzpicture}}
\par\end{centering}

\caption{Decoding latency of the BP decoder $L_{BP}$, the SCL decoder $L_{SCL}$
and the hybrid decoder $L_{hyb}$. Code parameters are $N=4096$,
$L=32$, $R=0.5$ and $i_{max}=50$. \label{fig:dec_latency}}
\end{figure}
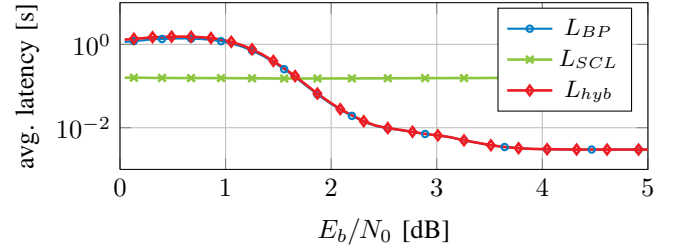
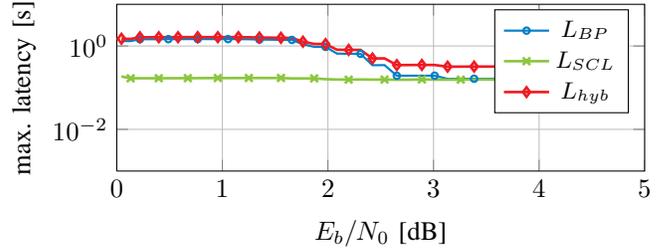

Whenever BP decoding fails, SCL decoding must be performed; then,
the total latency $L_{hyb}$ increases as shown in Fig.~\ref{fig:dec_latency}.
As the BP frame-error-rate is $\gamma_{BP,FER}\ll1$, the average
latency mainly depends on $L_{BP}.$ The BP decoding latency is strongly
related to the SNR due to the implemented early stopping mechanism.
In the low SNR region it is even above the SCL latency, which stems
from the fact that multiple BP codewords are decoded in parallel,
i.e., computation resources need to be shared. However, when compared
to SCL decoding, the average latency only increases slightly since
$L_{BP}\ll L_{SCL}$ for high SNR, i.e., only a few applied BP iterations.
When compared to other adaptive decoding concepts such as e.g., an
adaptive list size \cite{Li_adaptive_list}, our approach shows better
latency performance in the target SNR region.

\section{Conclusions}

In this work, we present a first SCL \textit{CUDA} implementation
for GPU computing and evaluate the achievable decoding throughputs.
As this algorithm turns out to be very challenging to speed-up, a
combination with BP decoding is considered. It is shown, that the
BP algorithm can be easily adapted for GPU computing. The decoding
throughput of the novel hybrid approach can be drastically increased,
while the average decoding latency decreases when compared to SCL
decoding. Further, in contrast to LDPC codes, almost no publications
regarding GPU-based polar decoding exist and we hope that this work
inspires other research groups to further investigate this challenging
problem.

\bibliographystyle{IEEEtran}
\bibliography{IEEEabrv,bibliography}

\begin{thebibliography}{10}
\providecommand{\url}[1]{#1}
\csname url@samestyle\endcsname
\providecommand{\newblock}{\relax}
\providecommand{\bibinfo}[2]{#2}
\providecommand{\BIBentrySTDinterwordspacing}{\spaceskip=0pt\relax}
\providecommand{\BIBentryALTinterwordstretchfactor}{4}
\providecommand{\BIBentryALTinterwordspacing}{\spaceskip=\fontdimen2\font plus
\BIBentryALTinterwordstretchfactor\fontdimen3\font minus
  \fontdimen4\font\relax}
\providecommand{\BIBforeignlanguage}[2]{{%
\expandafter\ifx\csname l@#1\endcsname\relax
\typeout{** WARNING: IEEEtran.bst: No hyphenation pattern has been}%
\typeout{** loaded for the language `#1'. Using the pattern for}%
\typeout{** the default language instead.}%
\else
\language=\csname l@#1\endcsname
\fi
#2}}
\providecommand{\BIBdecl}{\relax}
\BIBdecl

\bibitem{arikan_polar}
E.~{Arıkan}, ``Channel polarization: A method for constructing
  capacity-achieving codes for symmetric binary-input memoryless channels,''
  \emph{{IEEE} Trans. Inf. Theory}, vol.~55, no.~7, pp. 3051--3073, Jul. 2009.

\bibitem{arikan_challenges_coding}
E.~Arıkan, N.~ul~Hassan, M.~Lentmaier, G.~Montorsi, and J.~Sayir, ``Challenges
  and some new directions in channel coding,'' \emph{J. Commun. Netw.},
  vol.~17, no.~4, pp. 328--338, Aug. 2015.

\bibitem{tal/vardy_scl}
I.~Tal and A.~Vardy, ``List decoding of polar codes,'' \emph{{IEEE} Trans. Inf.
  Theory}, vol.~61, no.~5, pp. 2213--2226, May 2015.

\bibitem{niu_basic_concepts}
K.~Niu, K.~Chen, J.~Lin, and Q.~T. Zhang, ``Polar codes: Primary concepts and
  practical decoding algorithms,'' \emph{{IEEE} Commun. Mag.}, vol.~52, no.~7,
  pp. 192--203, Jul. 2014.

\bibitem{rost2015benefits}
P.~Rost, I.~Berberana, A.~Maeder, H.~Paul, V.~Suryaprakash, M.~Valenti,
  D.~W{\"u}bben, A.~Dekorsy, and G.~Fettweis, ``Benefits and challenges of
  virtualization in 5{G} radio access networks,'' \emph{{IEEE} Commun. Mag.},
  vol.~53, no.~12, pp. 75--82, 2015.

\bibitem{nikaein2015processing}
N.~Nikaein, ``Processing radio access network functions in the cloud: Critical
  issues and modeling,'' in \emph{Proceedings of the 6th International Workshop
  on Mobile Cloud Computing and Services}.\hskip 1em plus 0.5em minus
  0.4em\relax ACM, 2015, pp. 36--43.

\bibitem{nvidia_cuda}
NVIDIA, ``Performance guidelines,'' \emph{CUDA C Programming Guide}, Aug. 2014.

\bibitem{Falcao_ldpc_gpu}
G.~Falcao, L.~Sousa, and V.~Silva, ``Massively {LDPC} decoding on multicore
  architectures,'' \emph{{IEEE} Trans. Parallel Distrib. Syst.}, vol.~22,
  no.~2, pp. 309--322, Feb. 2011.

\bibitem{BP_arikan}
E.~{Arıkan}, ``A performance comparison of polar codes and reed-muller
  codes,'' \emph{IEEE Comm. Letters}, vol.~12, no.~6, Jun. 2008.

\bibitem{reddy_polar_bp_gpu}
{B. K. Reddy L. }and N.~Chandrachoodan, ``A {GPU} implementation of belief
  propagation decoder for polar codes,'' in \emph{Proc. Asilomar Conf. on
  Signals, Systems, and Computers}, Nov. 2012, pp. 1272--1276.

\bibitem{giard_gpu_decoder}
\BIBentryALTinterwordspacing
P.~Giard, G.~Sarkis, C.~Leroux, C.~Thibeault, and W.~J. Gross, ``Low-latency
  software polar decoders,'' \emph{CoRR}, vol. abs/1504.00353, 2015. [Online].
  Available: \url{http://arxiv.org/abs/1504.00353}
\BIBentrySTDinterwordspacing

\bibitem{yuan_hybrid_sc}
B.~Yuan and K.~K. Parhi, ``Algorithm and architecture for hybrid decoding of
  polar codes,'' in \emph{Proc. Asilomar Conf. on Signals, Systems, and
  Computers}, Nov. 2014, pp. 2050--2053.

\bibitem{Li_adaptive_list}
B.~Li, H.~Shen, and D.~Tse, ``An adaptive successive cancellation list decoder
  for polar codes with cyclic redundancy check,'' \emph{IEEE Comm. Letters},
  vol.~16, no.~12, pp. 2044--2047, Dec. 2012.

\bibitem{webdemo}
B.~Leible, M.~Stahl, and S.~Cammerer, ``{On-the-fly Polar Code Bit-Error-Rate
  Simulations},'' {Institute of Telecommunications, University of Stuttgart,
  Germany}, Oct. 2016, http://webdemo.inue.uni-stuttgart.de.

\bibitem{tal/vardy_construct_pcodes}
I.~Tal and A.~Vardy, ``How to construct polar codes,'' \emph{{IEEE} Trans. Inf.
  Theory}, vol.~59, no.~10, pp. 6562--6582, Oct. 2013.

\bibitem{peters2011fast}
H.~Peters, O.~Schulz-Hildebrandt, and N.~Luttenberger, ``Fast in-place,
  comparison-based sorting with {CUDA}: A study with bitonic sort,''
  \emph{Concurrency and Computation: Practice and Experience}, vol.~23, no.~7,
  pp. 681--693, 2011.

\bibitem{Li_decision_aided}
\BIBentryALTinterwordspacing
B.~Li, H.~Shen, and K.~Chen, ``A decision-aided parallel {SC}-list decoder for
  polar codes,'' \emph{CoRR}, vol. abs/1506.02955, 2015. [Online]. Available:
  \url{http://arxiv.org/abs/1506.02955}
\BIBentrySTDinterwordspacing

\bibitem{sung_bp_fpga}
Y.~S. Park, Y.~Tao, S.~Sun, and Z.~Zhang, ``A 4.68{Gb/s} belief propagation
  polar decoder with bit-splitting register file,'' in \emph{{Proc. IEEE Int.
  Symp. on VLSI}}, Jun. 2014, pp. 1--2.

\bibitem{yuan_stopping_cond}
B.~Yuan and K.~K. Parhi, ``Early stopping criteria for energy-efficient
  low-latency belief-propagation polar code decoders,'' \emph{{IEEE} Trans.
  Signal Process.}, vol.~62, no.~24, pp. 6496--6506, Dec. 2014.

\end{thebibliography}

\end{document}